\documentstyle[12pt]{article}

\begin{document}

\title{Dynamics of dissipative gravitational collapse}
\author{ L. Herrera$^{1,4}$\thanks{e-mail: laherrera@telcel.net.ve} and N. O. Santos$^{2,3,4}$\thanks
{e-mail: nos@cbpf.br and santos@ccr.jussieu.fr}\\
\small{$^1$Escuela de F\'{\i}sica, Facultad de Ciencias,} \\
\small{Universidad Central de Venezuela, Caracas, Venezuela.}\\
\small{$^2$Laborat\'orio Nacional de Computa\,{c}\~ao Cient\'{\i}fica,}\\
\small{25651-070 Petr\'opolis RJ, Brazil.}\\
\small{$^3$Centro Brasileiro de Pesquisas F\'{\i}sicas,}\\
\small{22290-180 Rio de Janeiro RJ, Brazil.}\\
{\small $^4$LERMA--UMR 8112--CNRS, ERGA Universit\'e PARIS VI}\\
{\small  4 place Jussieu, 75005 Paris Cedex 05, France.}\\}
\maketitle

\begin{abstract}
The Misner and Sharp approach to the study of gravitational collapse is extended to the dissipative case in, both, the streaming out and
the diffusion approximations. The role of different terms in the dynamical equation are analyzed in detail. The dynamical equation is then coupled to a causal transport equation in 
the context of Israel--Stewart theory.  The decreasing of the inertial mass density of the fluid, by a factor which depends  on its internal thermodynamics state, is reobtained, at any
time scale. In accordance with the equivalence principle, the same decreasing factor is obtained for the gravitational force term. Prospective applications of this result to some
astrophysical scenarios are 
 discussed.
\end{abstract}
\newpage
\section{Introduction}
Some years ago, Misner and Sharp \cite{MisnerSharp} and Misner \cite{Misner} provided a full account of the dynamical equations 
governing the adiabatic \cite{MisnerSharp}, and the
dissipative relativistic collapse in the streaming out approximation \cite{Misner}.

The relevance of dissipative processes in the study of gravitational collapse cannot be over emphasized. Indeed, dissipation
 due to the emission of massless
particles (photons and/or neutrinos) is a characteristic process in the
evolution of massive stars. In fact, it seems that the only plausible
mechanism to carry away the bulk of the binding energy of the collapsing
star, leading to a neutron star or black hole, is neutrino emission
\cite{1}. 

In the diffusion approximation, it is assumed that the energy flux of
radiation, as that of
thermal conduction, is proportional to the gradient of temperature. This
assumption is in general very sensible, since the mean free path of
particles responsible for the propagation of energy in stellar
interiors is in general very small as compared with the typical
length of the object.
Thus, for a main sequence star as the sun, the mean free path of
photons at the centre, is of the order of $2$ cm. Also, the
mean free path of trapped neutrinos in compact cores of densities
about $10^{12}$ g. cm$^{-3}$ becomes smaller than the size of the stellar
core \cite{3,4}.

Furthermore, the observational data collected from supernova 1987A
indicates that the regime of radiation transport prevailing during the
emission process, is closer to the diffusion approximation than to the
streaming out limit \cite{5}.

However in many other circumstances, the mean free path of particles
transporting energy may be large enough as to justify the  free streaming
approximation. Therefore it is advisable to include simultaneously both limiting  cases of radiative transport,
diffusion and streaming out, allowing for describing a wide range of
situations.

In a recent work \cite{6'} we have studied the effects of dissipation, in both limiting cases of radiative transport, within the context of quasi--static
approximation. This assumption is very sensible because the hydrostatic time scale is
very small for many phases of the life of a star. It is of the
order of 27 minutes for the sun, 4.5 seconds for a white dwarf and
$10^{-4}$ seconds for a neutron star of one solar mass and $10$km
radius \cite{7'}.
 However, during their evolution, self--gravitating objects may pass
through phases of intense dynamical activity, with time scales of the order
of magnitude of (or even smaller than) the hydrostatic time scale, and  for
which the quasi--static
approximation is clearly not reliable, e.g.,the collapse of very massive
stars \cite{8''}, and the quick collapse phase
preceding neutron star formation, see for example \cite{9'} and
references therein. In these cases it is mandatory to take into account
terms which describe departure from
equilibrium, i.e. a full dynamic description has to be used.

Thus our primary task here consists in extending the Misner dynamical equations as to include dissipation in the form of a radial heat flow (besides
  pure radiation). This  is presented in Section 3. Then in the following Section, the resulting dynamical equation is  coupled 
to the transport equation obtained in the context of the
M\"{u}ller--Israel--Stewart theory \cite{Muller67, IsSt76}.

After  doing that we show that the effective inertial mass density of a 
fluid element  reduces by a factor which depends on dissipative
variables. This result was already known (see \cite{eim} and references therein), but considered to be valid  only, just after leaving the
 equilibrium, on a time scale of the order of
relaxation time. The novelty here is, on the one hand, that such reduction of the effective inertial mass
 density is shown to be valid at an {\it
arbitrary time scale}, and on the other, that the``gravitational force'' term in the dynamical equation is also 
reduced by the same factor, as expected from the equivalence principle. Prospective applications of this result
to  astrophysical scenarios are discussed at the end. 

 In the next section the field equations, the conventions, and other useful formulae are introduced.
\section{The  energy--momentum tensor and the field equations}
In this section we  provide a full description of the matter distribution, the line element, both, inside and outside
  of the fluid boundary and the field equations this line element
must satisfy. Since we are going to follow closely the Misner approach \cite{Misner} we shall use comoving coordinates 
(for a description of gravitational collapse in non--comoving
coordinates, see \cite{H} and references therein).
\subsection{The interior spacetime}
We consider a spherically symmetric distribution of collapsing
fluid, which, for sake of completeness, we assume to be locally anisotropic,
undergoing dissipation in the form of heat flow and free streaming
radiation, bounded by a
spherical surface $\Sigma$. For such system the energy--momentum tensor is given by
\begin{equation}
T_{-}^{\alpha\beta}=(\mu+P_{\perp})V^{\alpha}V^{\beta}+P_{\perp}g^{\alpha\beta}+
(P_r-P_{\perp})\chi^{\alpha}\chi^{\beta}+q^{\alpha}V^{\beta}+V^{\alpha}q^{\beta}+
\epsilon l^{\alpha}l^{\beta}, \label{1}
\end{equation}
where, $\mu$ is the energy density, $P_r$ the radial pressure, $P_{\perp}$ is the tangential
pressure, $\epsilon$ is the radiation density, $V^{\alpha}$ is the four velocity of the fluid, $q^{\alpha}$ is the heat flux, $\chi^{\alpha}$ is a unit four vector along 
the radial direction and $l^{\alpha}$ is a null four vector. These quantities have to satisfy
\begin{equation}
V^{\alpha}V_{\alpha}=-1, \;\; V^{\alpha}q_{\alpha}=0, \;\; \chi^{\alpha}\chi_{\alpha}=1,
\;\; \chi^{\alpha}V_{\alpha}=0, \;\; l^{\alpha}l_{\alpha}=0. \label{2}
\end{equation}
We assume the interior metric to $\Sigma$ to be comoving, shear free for simplicity, and spherically symmetric, accordingly it may be written as
\begin{equation}
ds^2=-A^2(t,r)dt^2+B^2(t,r)(dr^2+r^2d\theta^2+r^2\sin^2\theta d\phi^2), \label{3}
\end{equation}
and hence
\begin{equation}
V^{\alpha}=A^{-1}\delta^{\alpha}_0, \;\; q^{\alpha}=q\delta^{\alpha}_1, \;\; 
l^{\alpha}=A^{-1}\delta^{\alpha}_0+B^{-1}\delta^{\alpha}_1, \;\; 
\chi^{\alpha}=B^{-1}\delta^{\alpha}_1, \label{4}
\end{equation}
where $q$ is a function of $t$ and $r$ and we have numbered the coordinates $x^0=t$, $x^1=r$,
$x^2=\theta$ and $x^3=\phi$.
Now the Einstein's field equations become with the help of (\ref{1}-\ref{4})
\begin{eqnarray}
8\pi T^-_{00}=8\pi(\mu+\epsilon)A^2=-\left(\frac{A}{B}\right)^2\left[2\frac{B^{\prime\prime}}{B}-\left(\frac{B^{\prime}}{B}\right)^2+\frac{4}{r}\frac{B^{\prime}}{B}\right]+
3\left(\frac{\dot{B}}{B}\right)^2, \label{5} \\
8\pi T^-_{01}=-8\pi (qB+\epsilon)AB=-2\left(\frac{\dot{B}^{\prime}}{B}-
\frac{B^{\prime}}{B}\frac{\dot{B}}{B}-\frac{A^{\prime}}{A}\frac{\dot{B}}{B}\right), \label{6} \\
8\pi T^-_{11}=8\pi (P_r+\epsilon)B^2=\left(\frac{B^{\prime}}{B}\right)^2+\frac{2}{r}\frac{B^{\prime}}{B}+
2\frac{A^{\prime}}{A}\frac{B^{\prime}}{B}+\frac{2}{r}\frac{A^{\prime}}{A} \nonumber \\
-\left(\frac{B}{A}\right)^2\left[2\frac{\ddot{B}}{B}+\left(\frac{\dot{B}}{B}\right)^2-
2\frac{\dot{A}}{A}\frac{\dot{B}}{B}\right],  \label{7} \\
8\pi T^-_{22}=\frac{8\pi T^-_{33}}{\sin^{2}\theta}=8\pi r^2P_{\perp} B^2=r^2\left[\frac{B^{\prime\prime}}{B}-\left(\frac{B^{\prime}}{B}\right)^2+
\frac{1}{r}\frac{B^{\prime}}{B}+\frac{A^{\prime\prime}}{A}+
\frac{1}{r}\frac{A^{\prime}}{A}\right] \nonumber \\
-r^2\left(\frac{B}{A}\right)^2\left[2\frac{\ddot{B}}{B}+
\left(\frac{\dot{B}}{B}\right)^2-2\frac{\dot{A}}{A}\frac{\dot{B}}{B}\right], \label{8}
\end{eqnarray}
where the dot and prime stand for differentiation with respect to $t$ and $r$.
The rate of expansion $\Theta={V^{\alpha}}_{;\alpha}$ of the fluid sphere is given, from (\ref{3}) and (\ref{4}), by
\begin{equation}
\Theta=3\frac{\dot{B}}{AB}, \label{8aa}
\end{equation}
and from (\ref{6}) we have
\begin{equation}
8\pi(qB+\epsilon)B=\frac{2}{3}\Theta^{\prime}. \label{8aaa}
\end{equation}
If $q>0$ and $\epsilon>0$, then from (\ref{8aaa}) we have $\Theta^{\prime}>0$ meaning that, if the system is collapsing $\Theta<0$, $q$ and/or $\epsilon$ decrease the rate of 
collapse towards the outer layers of matter. If $q=0$ and $\epsilon=0$ from (\ref{8aaa}) $\Theta^{\prime}=0$, which means that the collapse is homogeneous.

The mass function $m(t,r)$ of Cahill and McVittie \cite{Cahill} is obtained from the Riemann tensor component ${R_{23}}^{23}$ and is for metric (\ref{3})
\begin{equation}
m(t,r)=\frac{(rB)^3}{2}{R_{23}}^{23}=\frac{r^3}{2}\frac{B\dot{B}^2}{A^2}-\frac{r^3}{2}
\frac{B^{\prime 2}}{B}-r^2B^{\prime}. \label{8a}
\end{equation}

\subsection{The exterior spacetime}
The exterior spacetime to $\Sigma$ of the collapsing body is described by the outgoing Vaidya spacetime which models a radiating star and has metric
\begin{equation}
ds^2_+=-\left[1-\frac{2m(v)}{\rho}\right]dv^2-2dvd\rho+\rho^2(d\theta^2+\sin^2\theta d\phi^2),
\label{9}
\end{equation}
where $m$, the total mass inside $\Sigma$, is a function of the retarded time $v$. The surface $\Sigma$ described by the comoving coordinate system (\ref{3}) is $r=r_{\Sigma}=$ constant, while in the non comoving coordinate system (\ref{9}) is $\rho=\rho_{\Sigma}(v)$.
Matching the interior spacetime (\ref{3}) with source (\ref{1}) to the exterior spacetime (\ref{9}) by using Darmois junction conditions we obtain
\begin{eqnarray}
(P_r)_\Sigma=(qB)_\Sigma, \label{10} \\
(qB+\epsilon)_{\Sigma}=\frac{1}{4\pi}\left(\frac{L}{\rho^2}\right)_{\Sigma}, \label{11} \\
(rB)_{\Sigma}=\rho_{\Sigma}, \label{12} \\
\left(\frac{r^3}{2}\frac{B\dot{B}^2}{A^2}-\frac{r^3}{2}\frac{B^{\prime 2}}{B}-r^2B^{\prime}\right)_{\Sigma}=m(v),
\label{13} \\
A_{\Sigma}dt=\left(1-\frac{2m}{\rho}+2\frac{d\rho}{dv}\right)^{1/2}_{\Sigma} dv, \label{14}
\end{eqnarray}
where $L$ is defined as the total luminosity of the collapsing sphere as measured on its surface and is given by
\begin{equation}
L=L_{\infty}\left(1-\frac{2m}{\rho}+2\frac{d\rho}{dv}\right)^{-1}, \label{14a}
\end{equation}
and where
\begin{equation}
L_{\infty}=\frac{dm}{dv} \label{14b}
\end{equation}
is the total luminosity measured by an observer at rest at infinity. The result (\ref{10}) represents the continuity of the radial flux 
of momentum across $\Sigma$ in which only the heat flow $q$ appears. However, for the total radiation leaving $\Sigma$ (\ref{11}) the radiation $\epsilon$ contributes as well as $q$.
Although it might seem to be obvious, it is  perhaps important to stress that the radiation $\epsilon$ has the same null property associated to the exterior null radiation that
it produces, while the heat flux
$q$, producing the exterior null radiation too, is not a null flux. Relation (\ref{12}) is the equality of the proper radii as measured from the perimeter of the spherical surface
$\Sigma$ in both frames (\ref{3}) and (\ref{9}). The expression for the total mass (\ref{13}) is the corresponding mass function \cite{Cahill} given by (\ref{8}). The relationship of
proper times measured on $\Sigma$ with both frames (\ref{3}) and (\ref{14}) is given by (\ref{14}).
\section{Dynamical equations} 
For studying the dynamical properties of the field equations and following Misner and Sharp, let us  introduce the proper time derivative $D_t$ given by
\begin{equation}
D_t=\frac{1}{A}\frac{\partial}{\partial t}. \label{16} 
\end{equation}
Then using (\ref{16}) we can describe the velocity $U$ of the collapsing fluid as
\begin{equation}
U=rD_tB<0 \;\; (in \; the \;  case \; of \; collapse), \label{19}
\end{equation}
then (\ref{8a}) can be rewritten as
\begin{equation}
\frac{(rB)^{\prime}}{B}=\left[1+U^2-\frac{2m(t,r)}{rB}\right]^{1/2}=E. \label{20}
\end{equation}
The right hand side of (\ref{20}) can be interpreted as being the energy density $E$ of a collapsing fluid element. 
Next, by taking the proper time derivative of (\ref{8a}) we obtain
\begin{eqnarray}
D_tm=r^3\frac{B\dot{B}\ddot{B}}{A^3}+\frac{r^3}{2}\left(\frac{\dot{B}}{A}\right)^3-r^3
\frac{B\dot{A}\dot{B}^2}{A^4} \nonumber \\
+\frac{r^3}{2}\frac{\dot{B}B^{\prime 2}}{AB^2}
-r^3\frac{B^{\prime}\dot{B}^{\prime}}{AB}-r^2\frac{\dot{B}^{\prime}}{A}. \label{21a}
\end{eqnarray}
Considering (\ref{6}) and (\ref{7}) we can rewrite (\ref{21a}) as
\begin{equation}
D_tm=-4\pi\left[(P_r+\epsilon)r^3\frac{B^2\dot{B}}{A}+(qB+\epsilon)r^2B(rB)^{\prime}
\right], \label{22}
\end{equation}
and with (\ref{19}) and (\ref{20}) it becomes
\begin{equation}
D_tm=4\pi\left[-(P_r+\epsilon)U-(qB+\epsilon)E\right](rB)^2, \label{23}
\end{equation} 
which gives the rate of variation of the total energy inside a surface of radius $rB$. In the right hand side of (\ref{23}) $(P_r+\epsilon)|U|$ (in the case of collapse $U<0$) increases
the energy inside
$rB$ through the rate of work being done by $P_r$ and the induction field produced by $\epsilon$ and already observed by Misner \cite{Misner}. Clearly here the heat flux $q$ does not
appear since it does not produce an induction field. The second term $-(qB+\epsilon)E$ is the matter energy leaving the spherical surface.

 Another proper derivative that helps us to
study the dynamics of the collapsing system is the proper radial derivative $D_R$, where 
\begin{equation}
R=rB, \label{23aa}
\end{equation}
constructed from the radius of a spherical surface, as measured from its perimeter inside $\Sigma$, being
\begin{equation}
D_R=\frac{1}{R^{\prime}}\frac{\partial}{\partial r}. \label{23a}
\end{equation}
Then by taking the proper radial derivative (\ref{23a}) of (\ref{8a}) we obtain
\begin{eqnarray}
D_Rm=\frac{B}{(rB)^{\prime}}\left[-r^3\frac{B^{\prime}B^{\prime\prime}}{B^2}-r^2\frac{B^{\prime\prime}}{B}+\frac{r^3}{2}
\left(\frac{B^{\prime}}{B}\right)^3-\frac{3r^2}{2}\left(\frac{B^{\prime}}{B}\right)^2\right.
\nonumber \\
\left. -2r\frac{B^{\prime}}{B}-r^3\frac{A^{\prime}\dot{B}^2}{A^3}+\frac{r^3}{2}\frac{B^{\prime}
\dot{B}^2}{A^2B}+r^3\frac{\dot{B}\dot{B}^{\prime}}{A^2}+\frac{3r^2}{2}\left(\frac{\dot{B}}{A}
\right)^{2}\right]. \label{24}
\end{eqnarray}
Considering (\ref{5}) and (\ref{6}) then (\ref{24}) becomes
\begin{equation}
D_Rm=4\pi\left[\mu+\epsilon+(qB+\epsilon)\frac{rB\dot{B}}
{(rB)^{\prime}A}\right](rB)^2, \label{25}
\end{equation}
and with (\ref{19}) and (\ref{20}) we finally have
\begin{equation}
D_Rm=4\pi\left[\mu+\epsilon+(qB+\epsilon)\frac{U}{E}\right](rB)^2. \label{26}
\end{equation}
This expression gives the total energy entrapped between two neighboring spherical surfaces with respect to proper radius inside the fluid distribution. 
The first term on the right hand side of (\ref{26}) $\mu+\epsilon$ is due to the energy density plus the induction field and no heat flux appears. The second term $(qB+\epsilon)U/E$
is negative (in the case of collapse) and measures the out flux of heat and radiation.

Finally, we can obtain the acceleration $D_tU$ of a collapsing particle inside $\Sigma$. In order to do
that we start from (\ref{7}) and (\ref{8a}) which allows us to write
\begin{equation}
D_tU=-\left[m+4\pi(P_r+\epsilon)(rB)^3\right](rB)^{-2}+\frac{A^{\prime}}{A}
\frac{(rB)^{\prime}}{B^2}. \label{27}
\end{equation}
Calculating the $r$ component of the Bianchi identities, ${T_-^{1\beta}}_{;\beta}=0$, from (\ref{1}) we obtain
\begin{eqnarray}
P_r^{\prime}+\epsilon^{\prime}+(\mu+P_r+2\epsilon)\frac{A^{\prime}}{A}+2(\epsilon+
P_r-P_{\perp})\frac{(rB)^{\prime}}{rB} \nonumber \\
+(5qB+4\epsilon)\frac{\dot{B}}{A}+\dot{q}\frac{B^2}{A}+
\dot{\epsilon}\frac{B}{A}=0. \label{28}
\end{eqnarray}
Substituting the expression $A^{\prime}/A$ from (\ref{28}) into (\ref{27}) and considering (\ref{8a}), (\ref{16}), (\ref{23aa}) and (\ref{23a}) we obtain
\begin{eqnarray}
(\mu+P_r+2\epsilon)D_tU=-(\mu+P_r+2\epsilon)\left[m+4\pi(P_r+\epsilon)R^3\right]\frac{1}{R^2} \nonumber \\
-E^2
\left[D_R(P_r+\epsilon)+2\frac{\epsilon+P_r-P_{\perp}}{R}\right] \nonumber \\
-E\left[(5qB+4\epsilon)\frac{U}{R}+BD_tq+D_t\epsilon\right].
\label{29}
\end{eqnarray}  
Equation  (\ref{29}) has the  ``Newtonian'' form 
\begin{equation}
Force= Mass \; density \times Acceleration
\label{Newton}
\end{equation}

The first term with square brackets in the right hand side of (\ref{29}) represents the gravitational force. It shows that the gravitational force acting on a particle has a Newtonian
part with
$m$ and a purely relativistic gravitational contribution due to $P_r$ and $\epsilon$. The second term in square brackets represent the hydrodynamical forces. It consists of the usual 
pressure gradient term (including the contribution of the radiation to the pressure)
$D_R(P_r+\epsilon)<0$ counteracting collapse,
 and the anisotropic force term  $\epsilon+P_r-P_{\perp}$ which can be positive or negative thus, respectively,  accelerating more or counteracting the rate of collapse. In these two
terms the appearance of $\epsilon$ is due to the contribution  of radiation to the total energy density and radial pressure. The last term in square brackets contains the specific 
contribution of dissipation to the dynamics of the system. The first term within this bracket is positive ($U<0$) showing that the out flux of
$q>0$ and $\epsilon>0$ diminish the total energy inside the collapsing sphere thereby reducing the rate of collapse. It is interesting to observe the different effects that $q$ and
$\epsilon$ have on the dynamical behaviour of the collapsing fluid. The heat flux
$q$ helps only to slow down the rate of collapse by diminishing the energy inside the fluid sphere by producing an exterior outflowing radiation. On the other hand, the radiation
density
$\epsilon$ behaves not only in a similar way as $q$ by diminishing the energy of the collapsing sphere through the exterior outflow of radiation, but contributes too as an induction
field to the gravitational energy, first observed by Misner and collaborators \cite{Misner, Lindquist}, and contributes to the radial pressure $P_r$. The effects of $D_t\epsilon$ have
been discussed in detail in
\cite{Misner}. Thus it remains to analyse the effects of $D_tq$, this will be done in the next section after introducing the transport equation.

Before coming to the next section, we observe that from  (\ref{29})  the limit of hydrostatic
equilibrium when $U=0$, $q=0$ and $\epsilon=0$ can be achieved, producing
\begin{equation}
D_RP_R+\frac{2(P_R-P_{\perp})}{R}=-\frac{\mu+P_r}{R(R-2m)}
\left(m+4\pi P_rR^3\right),
\label{30}
\end{equation}
which is just the generalization of the TOV equation for anisotropic fluids \cite{Liang}, obtained in comoving coordinates in \cite{Chan} while studying dynamical instability
for radiating anisotropic collapse.
\section{Transport equation and its consequences}

As we mentioned before we shall use a transport equation derived from the M\"{u}ller-Israel-Stewart second
order phenomenological theory for dissipative fluids \cite{Muller67, IsSt76}.

Indeed, it is well known that the Maxwell-Fourier law for the radiation 
flux leads to a parabolic equation (diffusion equation) which predicts 
propagation of perturbation with infinite speed (see \cite{6}--\cite{8'} and 
references therein). This simple fact is at the origin of the pathologies 
\cite{9} found in the approaches of Eckart \cite{10} and Landau \cite{11} 
for relativistic dissipative processes.

\noindent
To overcome such difficulties, different relativistic 
theories with non-vanishing relaxation times have been proposed 
in the past \cite{Muller67, IsSt76, 14, 15}. The important point is that all these 
theories provide a heat transport equation which is not of 
Maxwell-Fourier type but of Cattaneo type \cite{18}, leading thereby to a 
hyperbolic equation for the propagation of thermal perturbation.
Thus the corresponding  transport equation for the heat flux reads \cite
{8} 
\begin{equation}
\tau
h^{\alpha\beta}V^{\gamma}q_{\beta;\gamma}+q^{\alpha}=-\kappa h^{\alpha\beta}
(T_{,\beta}+Ta_{\beta}) -\frac 12\kappa T^2\left( \frac{\tau
V^\beta }{\kappa T^2}\right) _{;\beta }q^\alpha ,  \label{21}
\end{equation}
where $h^{\mu \nu }$ is the projector onto the three space orthogonal to $%
V^\mu $,  
$\kappa $  denotes the thermal conductivity, and  $T$ and  $\tau$
denote temperature and relaxation time
respectively. 
Observe that, due to the symmetry of the problem, equation (\ref{21}) only has one independent component, which may be writtten as:
\begin{equation}
\tau(qB)\dot{}B+qAB^2=-\kappa (TA)^{\prime}-\frac{\kappa T^2 q B^2}{2}\left(\frac{\tau}{\kappa T^2}\right)\dot{}-\frac{3\tau \dot B
Bq}{2}. \label{V2} \end{equation}

Which, using (\ref{16}) and (\ref{27}), becomes

\begin{eqnarray}
BD_tq=-\frac{\kappa T}{\tau E}D_tU-\frac{\kappa T^{\prime}}{\tau B}-  
\frac{qB}{\tau}(1+\frac{\tau U}{R})-\nonumber \\
-\frac{\kappa
T}{\tau E}\left[m+4\pi(P_r+\epsilon)R^3\right]R^{-2}- 
\frac{\kappa T^2 q B}{2A\tau}\left(\frac{\tau}{\kappa T^2}\right)\dot{} -\frac{3U Bq}{2R}. \label{V3}
\end{eqnarray}

We can couple the transport equation in the form above (\ref{V3}) to the dynamical equation (\ref{29}), in order to bring out
the effects of dissipation (in the diffusion approximation) on the dynamics of the collapsing sphere. With that purpose,  let
us replace  (\ref{V3}) into (\ref{29}) (putting $\epsilon=0$), then we obtain after some rearrangements \begin{eqnarray}
(\mu+P_r)(1-\alpha)D_tU=F_{grav}(1-\alpha)+F_{hyd}+\nonumber \\
+\frac{E\kappa T^{\prime}}{\tau B}+\frac{EqB}{\tau}
+\frac{\kappa ET^2 q B}{2A \tau}\left(\frac{\tau}{\kappa T^2}\right)\dot{} -\frac{5U EBq}{2R}. \label{V4}
\end{eqnarray}
Where $F_{grav}$ and $F_{hyd}$ are defined by 
\begin{equation}
F_{grav}=-(\mu+P_r)\left[m+4\pi P_r R^3\right]\frac{1}{R^2}, \nonumber \\
\label{grav}
\end{equation}
and
\begin{equation}
F_{hyd}=
-E^2
\left[D_R P_r+2\frac{P_r-P_{\perp}}{R}\right],
\label{hyd}
\end{equation}
where $\alpha$ is given by
\begin{equation}
\alpha=\frac{\kappa T}{\tau (\mu+P_r)}.
\label{alpha}
\end{equation}

We can now analyze the overall effects of dissipation (in the diffussion approximation) on the evolution of the collapsing
sphere.

First of all observe that as $\alpha$ tends to $1$, the effective inertial mass density of the fluid element tends to zero.
This effect was known (see \cite{eim} and references therein), but only to be valid just after the system abandons the
equilibrium, on a time scale of the order of (or smaller than) the relaxation time. Here we see that it is present all along
the evolution. Furthermore we see that $F_{grav}$ is also multiplied by the factor $(1-\alpha)$. Indicating that the effective
gravitational attraction on any fluid element decreases by the same factor as the effective inertial mass (density). Which of
course is to be expected, from the equivalence principle. It is also worth mentioning that $F_{hyd}$ is in principle independent
(at least explicitly) on this factor.

Next observe that the third and the fourth terms, on the right hand side of
(\ref{V4}), are of oposite sign and of the same order of magnitude (at least in the case of not too strong gravitational
field). Finally, the sign and the order of magnitude of the fifth term is clearly, model dependent. Furthermore this term is absent in the ``truncated'' version of the theory (see
\cite{19}).

With these comments above in mind, let us imagine the following situation: A collapsing sphere evolves in such a way that the
value of $\alpha$ keeps increasing and  approaches the critical value of $1$. As this process takes place, the ensuing
decreasing of the gravitational force term would eventually lead to a change of the sign
of the right hand side of (\ref{V4}). Since that would happen for small  values of the
effective inertial mass density, that would imply a strong bouncing of the sphere, even for a small absolute value of the
right hand side of (\ref{V4}). At the origin of this effect, of course, is the equivalence principle, according to which the inertial mass and the passive gravitational mass are
equal. It is also worth noticing that changes
of inertial mass due to different physical phenomena, are familiar in other branches of physics. Thus
for example the effective inertial mass of an electron moving under a given force through
a crystal, differs from the value corresponding to an electron moving
under the same force in free space, and may even become negative (see \cite{kitel}).

For this picture to be physÁcally meaningful one should  first answer to the following questions:
\begin{itemize}
\item How close may $\alpha$ approach the critical value?
\item In what physical scenarios one could expect values of $\alpha$ close to the critical value?
\end{itemize}

Since these questions are related to each other, we shall answer to them simultaneously.

First of all it should be mentioned that from the analysis of stability and causality in
dissipative relativistic fluids \cite{9}, it follows that
causality and hyperbolicity (which imply stability) require for
dissipative viscous free systems
\begin{equation}
\label{tau1}\tau >\frac{\kappa T}{\mu +p}+ \frac{\kappa}{n}
\frac{c_s^2}{c_p},
\end{equation}
\begin{equation}
\label{tau2}\tau >\frac{\kappa T}{1-c_s^2} \left[ \frac 1{\mu
+p}+\frac {1}{nT} \left( \frac 1{c_v}-\frac{c_s^2}{c_p}\right) -%
\frac{2\alpha_p }{nc_v\kappa _T(\mu +p)}\right] 
\end{equation}
and 
\begin{equation}
\label{tau3}\tau >\frac \kappa {nc_s^2c_v}\left[ \frac{2\alpha_p T}{\kappa
_T(\mu +p)}-1\right],
\end{equation}
where $n$ and  $c_s$ denote the
particle density and the sound speed, $c_p$ and  $c_v$ are  the specific
heat at constant pressure and volume, $\kappa_T$ is the thermal expansion
coefficient  and $\alpha_p$  the isothermal compressibility.
These expressions are found from equations (146-148) in \cite{9},
taking the limit $\beta_o,\beta_2\rightarrow\infty$ and
$\alpha_i=0$ (this method was applied in \cite{8} to the
case in which only bulk viscous perturbations were present).
It should be kept in mind that the conditions above, are obtained within a linear
perturbative scheme.

Obviously, condition (\ref{tau1}) is violated at the critical point
(in fact it is violated, slightly below it).
However as we shall see below, it is not difficult to find physical
conditions for which the numerical values of variables entering in
the definition of $\alpha$ lead to $\alpha=1$. Therefore, the
relevant question is: Can a physical system actually reach the
critical point?  If the answer to this question is negative, then it
should be explained how a given system avoids the critical point.
Since, as mentioned before, numerical values of $\kappa$, $T$,
$\tau$, $\mu$, and $P_r$, leading to $\alpha\approx 1$ may
correspond to a non very exotic scenario. 
On the other hand, a positive answer seems to be prohibited by
causality and stability conditions. However, we shall conjecture here
that this might not be the case. In fact, the vanishing of the effective inertial
 mass density  at the
critical point, indicates that linear approximation is not valid at
that point. So it seems that the behaviour of the system close to the
critical point cannot be studied with a linear perturbative scheme, in which case
it might be possible for a given system to attain the critical point. Furthermore, in the general case (including viscosity) 
it may happen that causality breaks down beyond
the critical point \cite{justino}. Thus, it appears that there exist situations where a
given physical system may attain the critical point and even go beyond it.

Indeed, condition $\alpha \approx 1$ can be accomplished in non
very exotic systems. One of them is an interacting mixture of matter and
neutrinos, which is a well-known scenario during the formation of a neutron
star in a supernova explosion. In this case the heat conductivity
coefficient is given by \cite{Weinberg71,Shapiro89} 
\begin{equation}
\kappa =\frac 43bT^3\tau ,  \label{79d}
\end{equation}
where $\tau $ is the mean collision time and  $b=7N_\nu a/8$, with  $N_\nu $  the
number in neutrino flavors and $a$  the radiation constant. Assuming that
the two viscosity coefficients vanish, and since $p$ should be not larger than $\mu$, then 
\begin{equation}
\alpha =\frac{\kappa T}{\tau _\kappa \left( \mu +p\right) }\simeq \frac{%
\kappa T}{\tau \mu }.  \label{79e}
\end{equation}
Using usual units, the critical point is overtaken if 
\begin{equation}
T> \left(\frac{6\mu c^3}{7N_\nu a}\right)^{1/4}\sim 4.29\times 10^8\mu ^{1/4},
\label{79f}
\end{equation}
where we have adopted $\tau \sim \tau _\kappa $, $N_\nu =3$, $T$ is in
Kelvin and $\mu $ is given in g cm$^{-3}.$ The values of temperature, for
which $\alpha \sim 1$, are similar to the expected temperature that can be
reached during hot collapse in a supernova explosion \cite[$\S$ 18.6]{ShTe83}

Also, we may speculate that  $\alpha$ may
increase substantially (for non-negligible values of $\tau$) in a pre-supernovae event

Indeed, at the last stages of massive star evolution, the decreasing of the opacity
of the fluid, from very high values preventing the propagation of photons
and neutrinos (trapping \cite{3}), to smaller values, gives rise to
radiative heat conduction. Under these conditions both $\kappa$ and $T$
could be sufficiently large as to imply a substantial increase of
$\alpha$. Indeed, the values suggested in \cite{Ma} ($[\kappa] \approx
10^{37}$;
$[T] \approx 10^{13}$; $[\tau] \approx 10^{-4}$; $[\rho] \approx
10^{12}$, in c.g.s units and Kelvin ) lead to $\alpha \approx 1$. The obvious consequence of which
would be to enhance the efficiency of whatever expansion mechanism, of
the central core, at place, because of the decreasing of the inertial mass density and the gravitational force term.
At this point it is worth noticing that the relevance of relaxational effects on gravitational collapse has been recently exhibited and stressed (see
\cite{Collapse}, and references therein)

\section{Conclusions}
Following the scheme developped by Misner and Sharp, we have established the set of dynamical equations governing the evolution
 of collapsing dissipative spheres, taking into account,
both, the free--streaming and the difussion approximation. We have further coupled the dynamical equation with a heat transport
 equation obtained from the
M\"{u}ller-Israel-Stewart theory. The resulting equation brings out the relevance of a critical point ($\alpha=1$) for which, 
both the effective inertial mass density and the
gravitational frorce term vanish. We have shown that in principle that critical point may be attained under acceptable physical
 conditions (e.g. a supernova scenario) and have speculated about the possibility
that in that case, a collapsing sphere bounces. Of course the eventual application of this model to a supernova, would require 
much more details about the astrophysical settings.

Before concluding we would like to make the following remarks:
\begin{enumerate}
\item It should be noticed that the appearance of the factor $1-\alpha$ in the inertial mass density and the gravitational force
term, is related to the first term on the left of equation (\ref{21}). But this is the term which introduces causality in the
transport equation and therefore, is to be expected in any causal theory of dissipation. Accordingly our main result is also
expected to hold for a general family of theories which includes the  M\"{u}ller-Israel-Stewart theory. However, whereas the
mere  appearance of the factor $1-\alpha$ in the dynamical equation is a very general result, the possibility of reaching the
critical point and the physical consequences derived from that, will depend on the specific theory of dissipation to be
adopted.
In this same line of arguments, it should be noticed that the transport equations in the M\"{u}ller-Israel-Stewart theory coincide with those obtained from the kinetic theory.
Accordingly the critical point should also appear in a kinetic theory approach.
\item Observe the formal similarity between the critical point and 
the equation of state for an inflationary scenario ($\mu = - P_r$) without dissipation. This kind of equation of state has been
recently proposed to describe the interior of a cold compact object without event horizons, and which would represent an
alternative to black holes \cite{mazur}. One could speculate with the possibility that such interior could be described
instead, by a dissipative fluid with $\alpha$ tending to $1$.
\item It should be clear that the analysis presented here depends strictly 
on the validity of the diffusion approximation, which in turn depends on 
the assumption of local thermodynamical equilibrium (LTE). Therefore, only 
small deviations from LTE can be considered in the context of this work. Thus when we state that the effective inertial
 mass density  decreases by the factor $(1-\alpha)$, at all time
scales, this means at time scales within which the system is not very far from LTE.
\item For the sake of completeness we have considered an anisotropic fluid 
(instead of an isotropic one, $P_r=P_\perp$), leaving the origin of such anisotropy completely 
unspecified. As it is apparent, anisotropy does not affect the most important 
result obtained here (e.g. the existence of the critical point). However, should anisotropy be related 
to viscosity, then for consistency the anisotropic pressure tensor should be 
subjected to the Israel-Stewart causal evolution equation for shear viscosity.
\end{enumerate}

\end{document}